\documentstyle[11pt,newpasp,twoside,psfig]{article}
\begin{document}

\title{The microarcsecond quasar J1819+3845}

\author{J. Dennett-Thorpe$^{1}$ \& A. G. de Bruyn$^{2,1}$}

\affil{(1) Kapteyn Institute, Postbus 800, Groningen, The Netherlands;
(2) ASTRON, Postbus 2, Dwingeloo, The Netherlands }

\begin{abstract}
We present new WSRT observations of the micro-arcsecond quasar
J1819+3845. All short term variations are attributed to interstellar
scintillation of a source which is at most 30 micro-arcseconds in
diameter. The timescale of the modulations changes
over the year, which we interpret as due to a peculiar velocity of the
scattering medium. The scintillation behaviour can be used
to determine sub-structure in the source. 
\end{abstract}


\section{ Introduction }

The quasar J1819+3845 is, at radio wavelengths, the most extremely
variable extragalactic source known (Dennett-Thorpe \& de Bruyn, 2000 ApJL 529
65). At the beginning of 1999 it showed peak to peak variations of
more than 600\% at 5\,GHz, with just an hour between the minima.
Intrinsic variations of this rapidity would require not only absurd
brightness temperatures, but a source which would be small enough to
scintillate due to the Galactic interstellar medium. We therefore
consider refractive scintillations as the only cause of the hourly
variations and concluded that: \\ {
$\bullet$} the scattering occurs in a nearby ($\sim$25pc) screen\\
{ $\bullet$} $>$55\% of the flux density at 5\,GHz must be
contained within a diameter of 30\,microarcseconds: or 5 light-months
at the source redshift ($z$=0.53). \\ { $\bullet$} If the
scintillations are due to one component, the brightness temperature
T$_B$=5\,10$^{12}$K\\ This high T$_B$ combined with a spectral peak
$\sim$ 100GHz, requires that the source is a transient phenomenon ($<$
5 months); has an exotic emission process; or has Doppler factors of
$\sim$15 and a continuous energy input.

 We conducted a monitoring campaign with the WSRT to address the
velocity of the scattering plasma (a critical unknown in the
application of the scintillation theory), the longevity of the source
and the source structure.


\section{ Seasonal variations }
From the monitoring campaign we found that:\\
{$\bullet$} The source continued to scintillate
throughout the 14 month observing  period.\\ 
{$\bullet$} There is a yearly variation in the timescale of the
modulations over this period; with the fastest variations around
February, and the slowest in August. The variations in August are
around an order of magnitude slower than those in February.\\ 
{$\bullet$} There is no indication for a change in the strength of the
modulations.\\ These results most easily
explained as due to a seasonal change in the relative velocity of the
Earth and the scattering plasma. This changes the speed at which we
move through the projected scintillation pattern. As J1819+3845 is
very close to the ecliptic pole, there is very little change in the
timescale of the modulations due to the Earth's motion. We conclude,
from fits to the observed timescale (measured in several different
manners), that {\it the scattering plasma has a velocity $\sim$ 25
km\,s$^{-1}$ w.r.t. the LSR}. This velocity is in
the opposite sense expected due to differential Galactic rotation, and
further strengthens the notion that the {\it scattering occurs
predominantly in a discrete structure}, not throughout the ISM.

\begin{figure}
\psfig{file=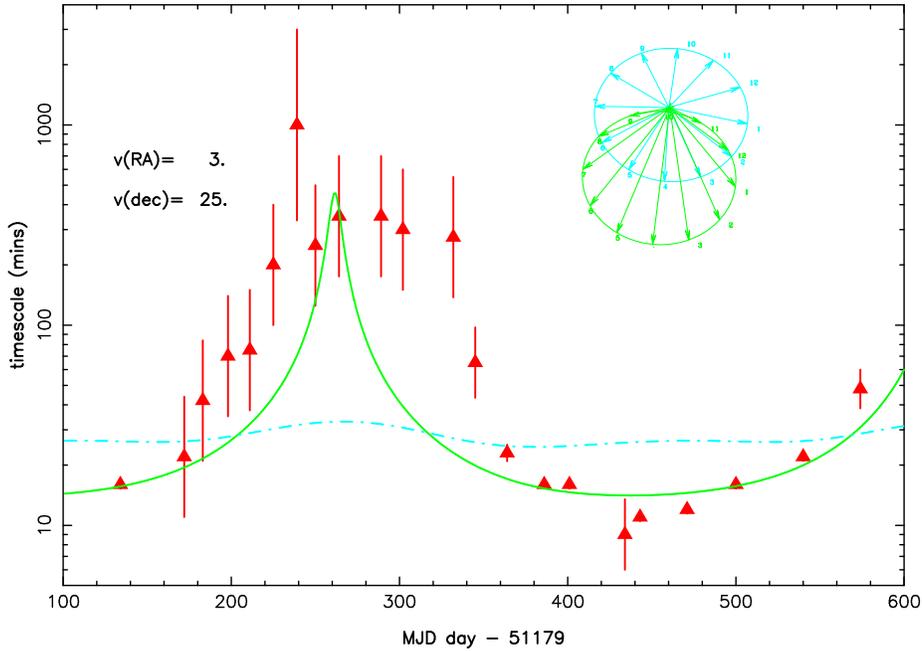,width=14cm,angle=-90}
\caption{The timescale variations throughout the
year (dotted line = no plasma velocity, solid line = best
fit). Error bars shown on {\bf all} data. Note logarithmic vertical scale.
Inset shows the relative Earth--plasma velocity vectors throughout
the year with and without plasma velocity.}
\end{figure}

The scintillations from a finite source are weighted towards the
medium where the refractive scale (medium dependent), Fresnel size
(distance dependent) and source size are matched. Thus
 quasars (unlike pulsars) will be more affected by
nearby scattering material.

We note that the character of the modulations appears different around
May (smaller, faster variations accompanied with larger excursions)
than at other times of the year (where the light curves seem
smoother). Furthermore, the period of slow modulations, which extends
for almost 6 months, is longer than expected if the change in
timescale is solely due to the peculair velocity of the plasma. Work
is underway to investigate that, in addition to the peculiar velocity
of the plasma, there may be additional affects due to source
elongation, anisotropic turbulence or contributions from other
scattering media (including the extended Galactic ISM).


\section{ Implications for the AGN}

\indent{$\bullet$}We have now measured the effective velocity of the
scattering medium, and have therefore eliminated a previous unknown in
our calculation. Assuming that all the flux density in the
`scintillating component' (75\,mJy at 5GHz) is in a single component,
we have a firm limit T$_B >$5\,10$^{12}$K.

{$\bullet$} There is evidence for source structure. As the
modulations do not reach 100\% at any frequency, the source must have
a component which is larger than the scattering zone, or $>$
30$\mu$arcsecs at 5GHz.

Furthermore, there is evidence for substructure in the source, on
scales $\sim$ the scattering zone, from the persistent asymmetries in
the lightcurves. Variations in polarised flux also support multiple
scintillating components.

{$\bullet$}Doppler factors $\cal D \sim$ 15 would allow for the Compton losses
which limit the intrinsic T$_B$. (If $\cal D$ does not greatly exceed
this, the source should be Compton limited and therefore a copious X-ray
emitter.) Even so, the spectral peak $\sim$ 100\,GHz means that these
electrons would lose their energy in a matter of weeks.

{$\bullet$}The source shows intrinsic variations in polarisation over
a period of months, but at centimeter wavelengths, the intrinsic
luminosity of the source has been remarkably constant. At most it has
shown a 20\% increase.

{\sf How the source remains so small, so hot, and also so constant is a puzzle.}
\end{document}